\documentclass[letterpaper]{jpconf}

\usepackage{graphicx}

\begin{document}

\title{Performance of the CMS tracking detectors from the 2009 LHC run}

\author{Keith A. Ulmer, on behalf of the CMS collaboration}

\address{University of Colorado}

\begin{abstract}
The 2009 run provided the first proton-proton collisions from the Large Hadron Collider (LHC) at center of
mass energies of 900 GeV and 2.36 TeV. The Compact Muon Solenoid (CMS) experiment has recorded a large 
sample of minimum bias events from these collisions. We present results from the all silicon tracking 
detectors from this run. The performance of the tracker and track reconstruction algorithms are considered 
including signal-to-noise, efficiencies and comparisons to simulation for track parameter and resonance
reconstruction performance.
\end{abstract}

\section{Introduction}
The first collisions at CMS were recorded in December 2009 at energies
of $\sqrt{s}~=~900$~GeV and 2.36~TeV.  The reconstructed tracks in this
data have been studied extensively to commission the tracking detectors
and to reconstruct basic physics objects as a demonstration of the
performance. In these proceedings, we describe the results of some of
these studies.

The CMS tracker is an all silicon design consisting
of two main detectors: a silicon pixel detector, covering the region
from $\approx$~4~cm to $\approx$11~cm in radius, and $\approx$~27~cm
on either side of the collision point along the LHC beam axis, and a
silicon strip detector, covering the region from $\approx$~25~cm to
$\approx$110~cm in radius, and $\approx$~280~cm on either side of the
collision point along the LHC beam axis. The pixel detector has 66
million channels covering an active surface area of about 1 m$^2$ that
provide precise three-dimensional information about charged
track positions.
The strip detector has 9.3 million channels covering an area of
198 m$^2$. Both are arranged in a pattern of concentric cylinders
to cover the central region
and discs to cover the forward regions. Additional details can be
found in~\cite{CMS}.

\section{Tracker Performance}
The 2009 run was used to evaluate the performance of the tracker
and to perform various calibrations needed to operate with the 
best performance and highest efficiency possible in future runs. A
fraction of 98.4\% of the pixel channels were operational. The 
remaining 1.6\% of inoperable channels was due mainly to channels
with a slow signal rise time, which may be recoverable for future 
running. The strip detector operated with 97.2\% of its channels
functioning well, with most of the inoperable channels due to
voltage supply shorts or leaks in the cooling system. The system
is built with large redundancy and the effect of the missing
channels on track reconstruction performance is minimal.

The 2009 LHC collisions provided the first opportunity to record
tracks from particles produced in time with the LHC clock.
The pixel detector 
readout system uses the 40~MHz LHC clock as input.  Signals from 
the CMS trigger system must arrive at the correct time within the 
25~ns clock cycle to associate the correct bunch crossing time stamp 
with any signal above the readout threshold.  An optimally phased
clock signal will maximize the number of pixels observed in the 
recorded clusters.  The overall trigger timing was adjusted by varying 
the clock phase until the average cluster size was maximized.
A finer module-by-module adjustment of the clock phase will be 
performed when higher trigger rates become available.
A similar timing scan was performed for the strip detector and 
the delay with the largest mean collected charge per cluster
was chosen
for future running. The timing for the strip detector is not as
sensitive as for the pixels due to the wider pulse shape.

\begin{figure}[hbtp]
  \begin{center}
    \resizebox{0.50\textwidth}{!}{\includegraphics{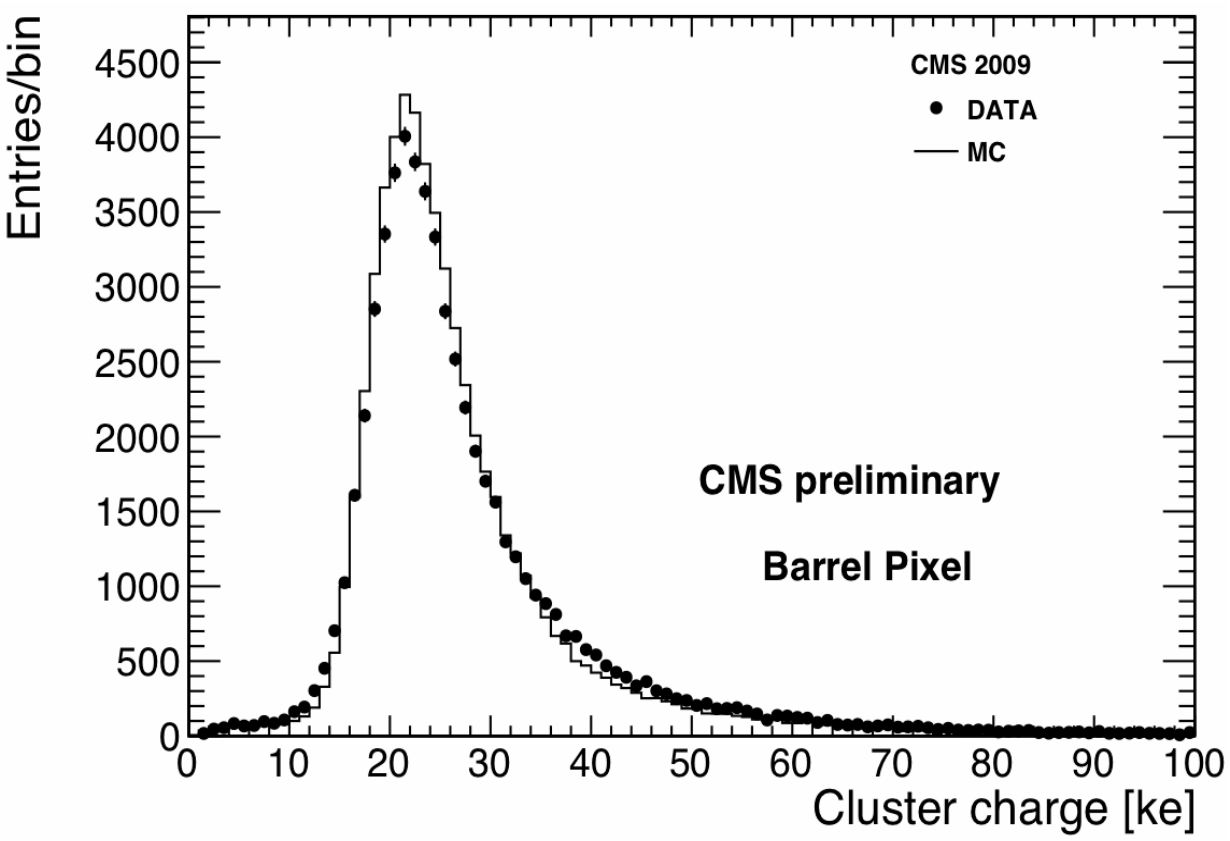}}
    \resizebox{0.38\textwidth}{!}{\includegraphics{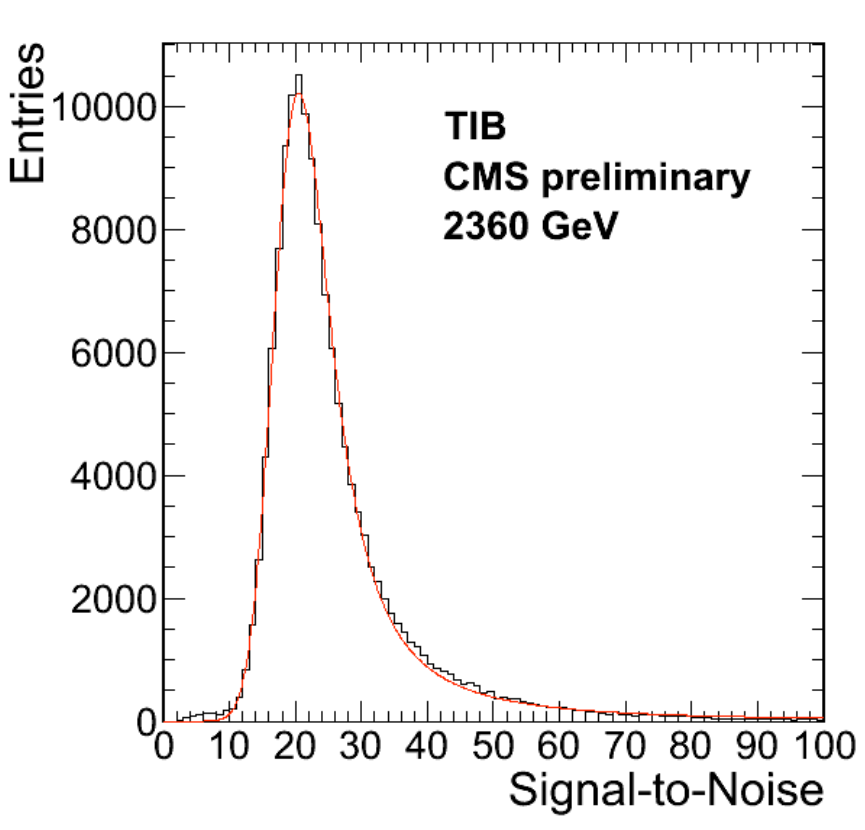}}
    \caption{Charge per cluster in the pixel barrel layers compared to simulation (left) and 
    strip signal/noise ratio for clusters on tracks in inner barrel layers (right).}
    \label{fig:det}
  \end{center}
\end{figure}

After optimizing the timing delays, the performance of the pixel 
and strip detectors was evaluated. The left plot of
Figure~\ref{fig:det} shows the
distribution of collected charge per cluster recorded by the barrel
pixel detectors compared to simulation, where in both data and simulation
the charge is normalized to the track path length through the silicon.
The peak of the charge collection shows good agreement with simulation,
though the data distribution is somewhat more broad than simulation.

The strip detector signal over noise was measured for each module where 
the signal was taken from clusters reconstructed in tracks, and noise
was evaluated in calibration runs. The S/N varies with the thickness and
length of the sensor. A representative plot is shown for inner strip barrel
sensors (TIB) in the right plot of Figure~\ref{fig:det}, where the
characteristic Landau distribution of collected charge is observed. The
most probable values of the S/N values range from 18.8 to 24.5 depending
on the sensor thickness and are in good agreement with simulation. 
The efficiency of the strip detector sensors was measured by using
reconstructed tracks to search for the presence of a recorded hit
on modules known to have been traversed by the track. For operational
modules the efficiency is measured to be in excess of $99.9\%$.

\section{Track Reconstruction Performance}

The standard track reconstruction at CMS is performed by the
combinatorial track finder~\cite{CMS_NOTE_2006_041}.  
Tracks are seeded from either triplets of hits in the tracker 
or pairs of hits with an additional constraint from the
beamspot or a pixel vertex, yielding an initial estimate of the 
trajectory, including its uncertainty.  The seed is then propagated 
outward in a search for compatible hits.  As hits are found, they are 
added to the trajectory and the track parameters and uncertainties are 
updated.  This search continues until either the boundary of the tracker is
reached or no more compatible hits can be found.

\begin{figure}[hbtp]
  \begin{center}
    \resizebox{0.41\textwidth}{!}{\includegraphics{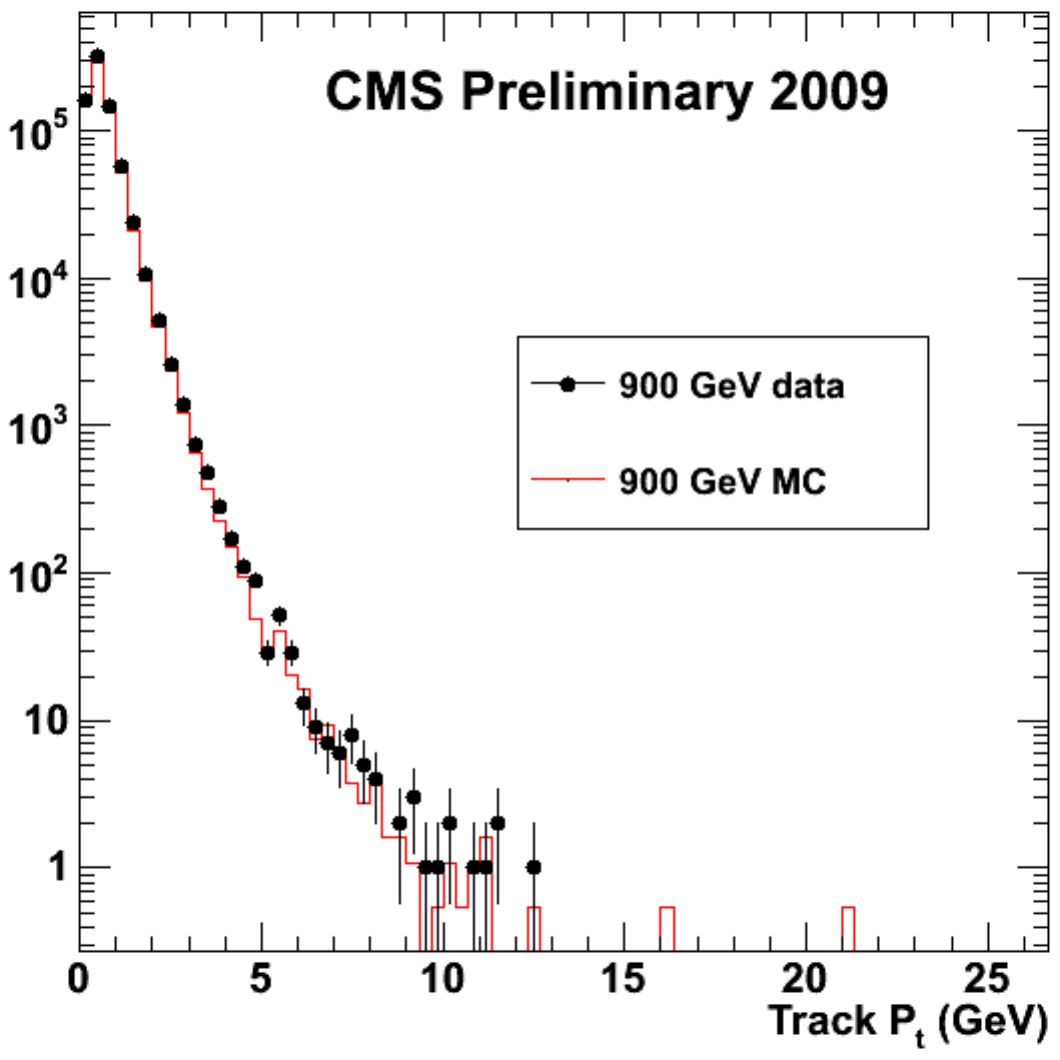}}
    \hspace{0.05\linewidth}
    \resizebox{0.45\textwidth}{!}{\includegraphics{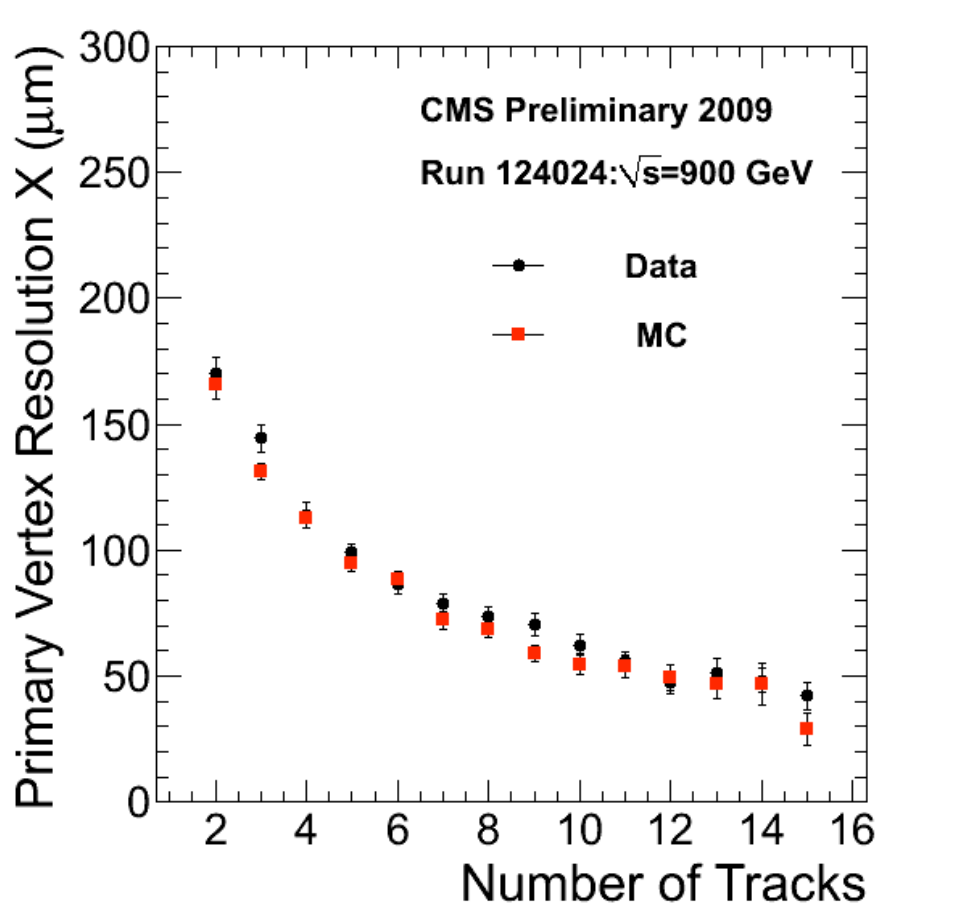}}
    \caption{Track transverse momentum distribution compared to
    simulation (left), and primary vertex resolution vs number of tracks
    compared to simulation (right).}
    \label{fig:track}
  \end{center}
\end{figure}

Reconstructed tracks
are filtered to remove tracks that are likely fakes and to provide a means of quantifying
the quality of the remaining tracks.  The filtering uses information on the number of
hits, the normalized $\chi^2$ of the track, and the compatibility of the track originating
from a pixel vertex.  Tracks that pass the tightest selection are labeled
\textit{highPurity}~\cite{CMS_PAS_TRK_10_001}.
The \textit{highPurity} tracks are selected, with additional requirements of
$|d_z| < 10\sigma_z$ (where $\sigma_z$ is the combined track and 
primary vertex uncertainty)
and  $\sigma_{p_T}/p_T < 10\%$ to compare
the data and simulation. 
The left plot of Figure~\ref{fig:track} shows the results of this comparison
for the track transverse momentum, where the
distribution has been normalized to
the number of reconstructed tracks in the data. As with $p_T$, there is general agreement
between the data and simulation in the shape of all variables.

The reconstruction of the primary interaction vertex in the event starts from the track
collection.  After filtering the tracks, they are clustered in $z$ and the cluster is fit
with an adaptive vertex fit, where tracks in the vertex are assigned a weight between 0
and 1 based on their proximity to the common vertex.
The primary vertex resolution is strongly dependant on the number of tracks used in fitting
the vertex.  To measure the resolution,
the tracks in an event are split into two different sets
and used to independently fit the primary vertex.  The distribution of the difference in
the fitted vertex positions can then be used to extract the resolution by fitting a
Gaussian to the distribution and dividing $\sigma$ by $\sqrt{2}$.
The right plot of Figure~\ref{fig:track} shows the $x$ resolution of the primary vertex
position as a function of the number of tracks in the vertex. Good agreement between
data and simulation is observed.

\begin{figure}[hbtp]
  \begin{center}
    \resizebox{0.32\textwidth}{!}{\includegraphics{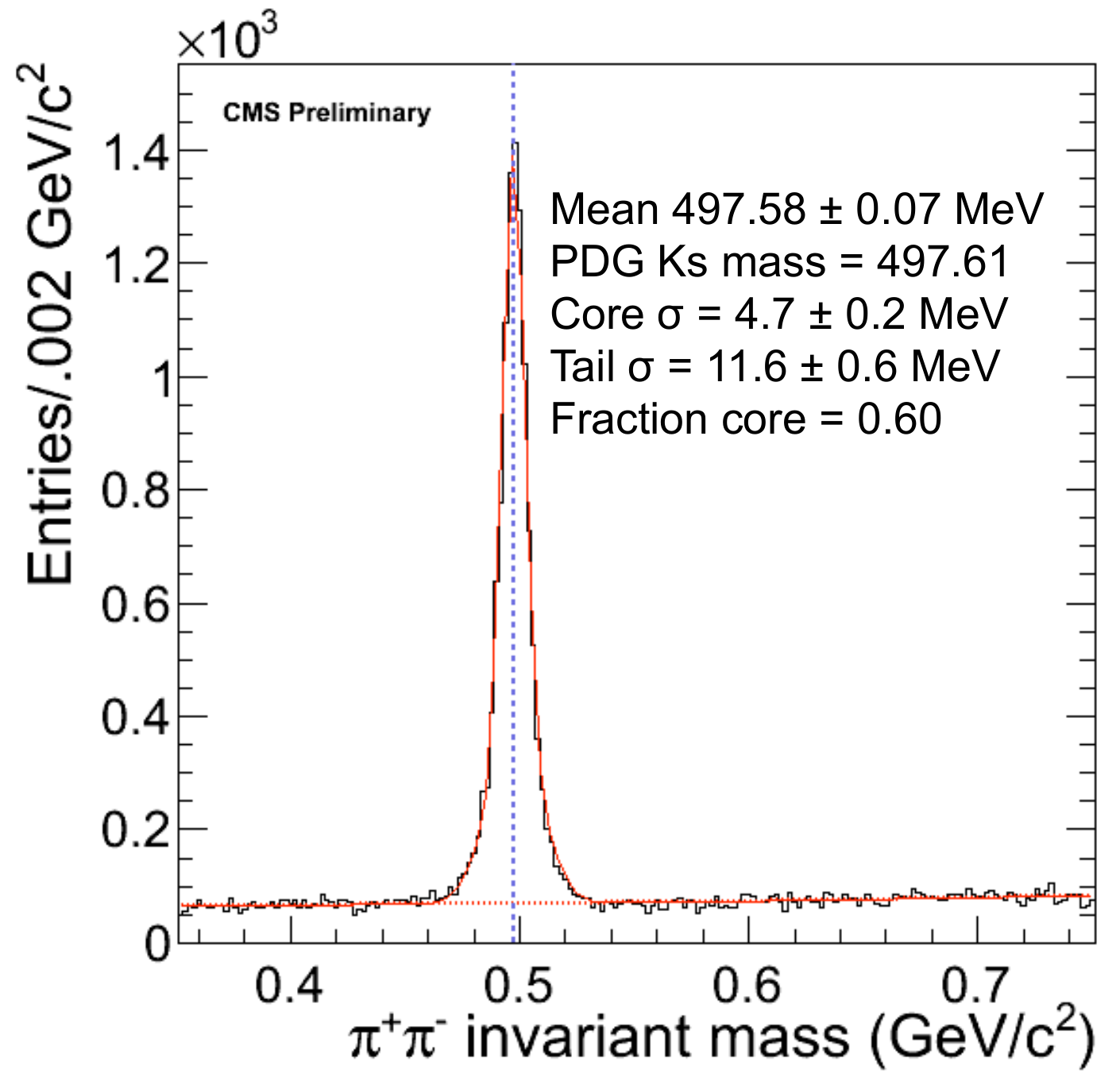}}
    \resizebox{0.32\textwidth}{!}{\includegraphics{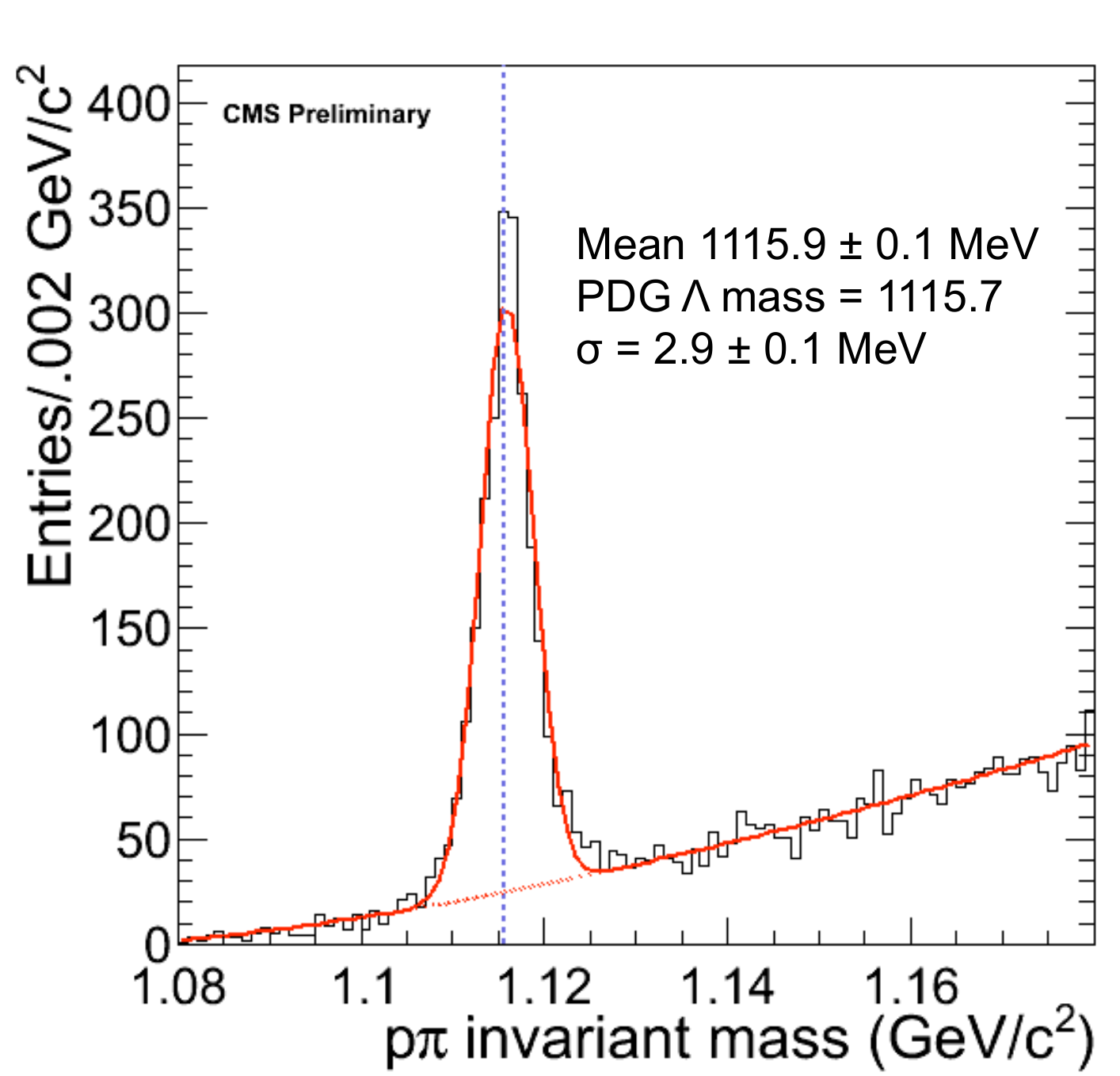}}
    \resizebox{0.31\textwidth}{!}{\includegraphics{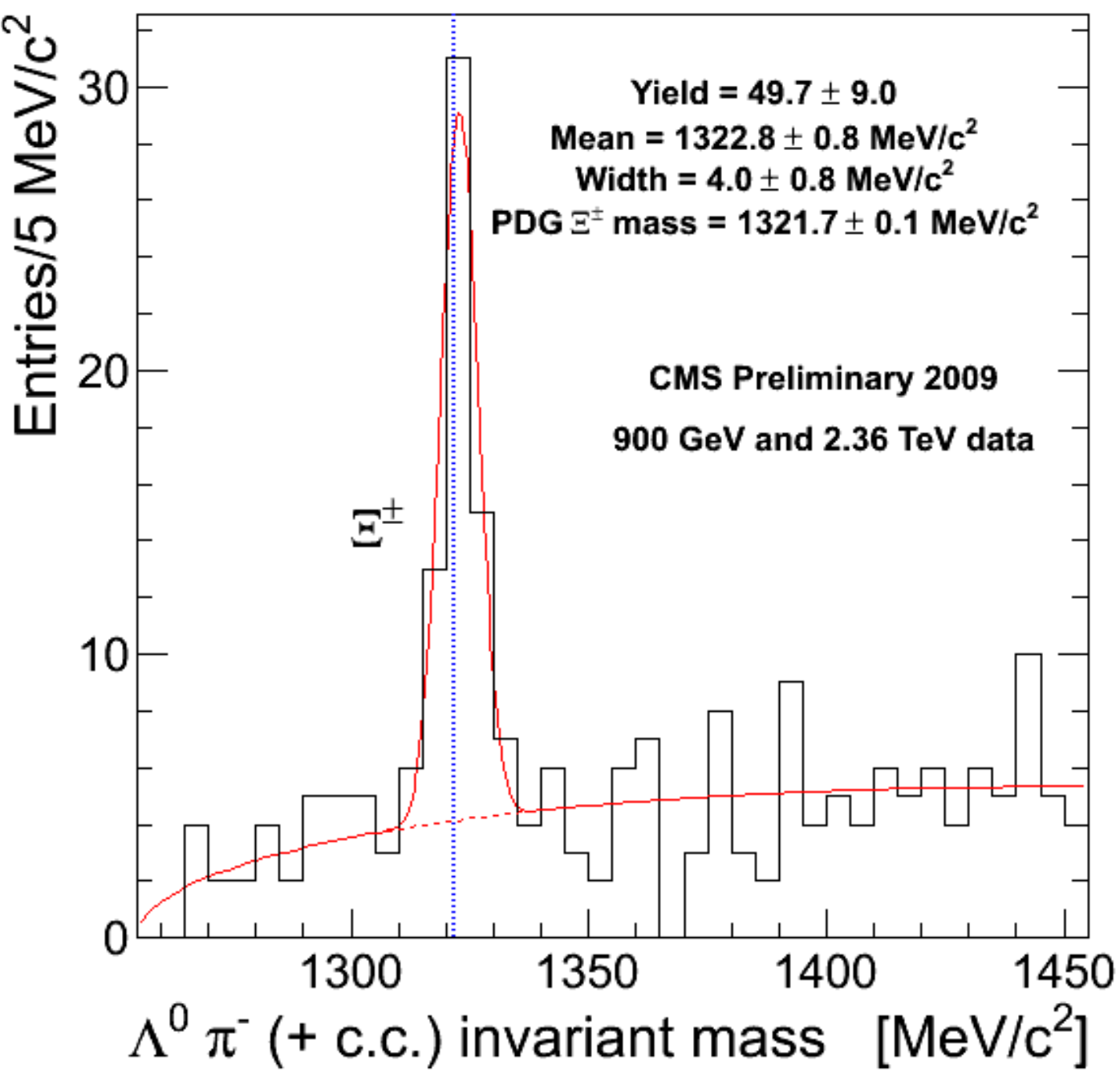}}
    \caption{Invariant mass distributions from $\pi^+\pi^-$, $p\pi^-$ and $\Lambda^0\pi^-$
    combinations (left to right).}
    \label{fig:res}
  \end{center}
\end{figure}

With a collection of well reconstructed tracks, a number of resonances can be
reconstructed from our data sample. Long-lived strange particles, $K_S^0$ and
$\Lambda^0$, were reconstructed in their decays to two charged particles, 
$\pi^+\pi^-$ and $p\pi^-$, respectively. 
\footnote{Charge conjugate decays are implied throughout.}
The reconstruction is performed by finding oppositely charged tracks which are
detached from the primary vertex and form a good vertex with an appropriate 
invariant mass. The tracks are required to have at least 6 hits, a normalized
$\chi^2 < 5$ and a transverse impact parameter with respect to the beamspot
greater than $0.5\sigma$. The reconstructed decay vertex must have a 
$\chi^2$ less than 7 and a transverse separation from the beamspot greater 
than 15$\sigma$. The resulting distributions show clear peaks
for these resonances as shown in Figure~\ref{fig:res}.

With the sample of $\Lambda^0$ particles, the cascade decay 
$\Xi^-\rightarrow\Lambda^0\pi^-$ can also be reconstructed. The $\pi^-$ from the
$\Xi^-$ will also be detached from the primary vertex due to the $\Xi^-$ lifetime.
$\Lambda^0$ candidates are reconstructed as described above, except with a looser
transverse significance cut of 10, and combined with charged tracks with the
same sign as the pion in the $\Lambda^0$ decay. The $\Lambda^0\pi^-$ fit applies a
$\Lambda^0$ mass constraint and the vertex is required to have a fit probability greater
than 1\%. All three tracks involved in the decay were required to have at least 6 valid
hits and a 3D impact parameter with respect to the primary vertex greater than 3$\sigma$.
The resulting mass peak is shown in Figure~\ref{fig:res}.

\section{Conclusion}
The performance of the CMS tracker has been studied using the collision data provided by
the LHC in the 2009 run. The tracker has been shown to perform well and is in good agreement
with the expectations from simulation. The track reconstruction algorithm has also been shown
to robustly reconstruct charged tracks and allow for the reconstruction of the $K_S^0$, $\Lambda^0$
and $\Xi^-$ resonances.


\begin{thebibliography}{0}
\bibitem{CMS}
The CMS Collaboration, JINST\ {\bf 0803}, S08004 (2008).

\bibitem{CMS_NOTE_2006_041}
Adam, W., Mangano, B., Speer, Th. and Todorov, T., CMS Note {\bf 2006/041}, 
\begin{verbatim} http://cms.cern.ch/iCMS/jsp/openfile.jsp?type=NOTE&year=2006&files=NOTE2006_041.pdf\end{verbatim} (2006).

\bibitem{CMS_PAS_TRK_10_001}
The CMS Collaboration, CMS Physics Analysis Summary {\bf TRK-10-001},
\begin{verbatim} http://cms-physics.web.cern.ch/cms-physics/public/TRK-10-001-pas.pdf\end{verbatim} (2010).

\end{thebibliography}
\end{document}